\documentclass{article}

\usepackage{graphicx}
\begin{document}

\title{Linear energy divergences in Coulomb gauge QCD}

\author{A. Andra\v si  \footnote{aandrasi@rudjer.irb.hr} \\
{\it 'Rudjer Bo\v skovi\'c' Institute, Zagreb, Croatia} }

\date{01 Feb. 2011}

\maketitle

\begin{abstract}

{\noindent The structure of linear energy divergences is analysed on the example
 of one graph to 3-loop order. Such dangerous divergences do cancel when all graphs
 are added, but next to leading divergences do not cancel out.}\\

\noindent{Pacs numbers: 11.15.Bt; 11.10.Gh} \\

\end{abstract}

\vfill\newpage

\section{Introduction}

The Coulomb gauge in non-Abelian gauge theories is a very good example of a physical gauge.
It is manifestly unitary. Although there are ghosts, their propagators have no poles.
The propagators are closely related to the polarization states of real spin-1 particles.
Nevertheless there are problems concerned with energy divergences \cite{JCTaylor}.
In individual Feynman graphs there appear even linear energy divergences. These are
divergences over the energy integration in a loop, for fixed values of the 3-momentum,
of the form

\begin{equation}
\label{1}
\int dk_0 F
\end{equation}
where $F$ is independent of $k_0$. They do cancel when all graphs are combined \cite{Mohapatra}.
However, it makes one uneasy in manipulating divergent and unregulated integrals.

\section{The graph $ 2B(b, 0i0)$}

We have studied the renormalization in Coulomb gauge QCD to three-loop order in Hamiltonian
formalism \cite{AA&JCT ANN2}. It was shown that to three loops the UV divergences cannot be
consistently absorbed by the Christ-Lee term \cite{Christ-Lee}. In this paper we show in detail
how dangerous the linear divergences are on the example of one graph with fermion loop
three-point function. The graph is shown in fig.1.
\begin{figure}[h]
\begin{center}
\includegraphics[width=7.5cm]{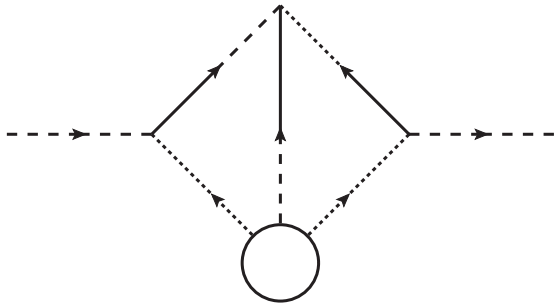}%
\caption{Graph $ 2B(b, 0i0)$ which is an example of the graph containing linear energy divergences}
\end{center}
\end{figure}

We use the same notation and graphical conventions as in \cite{AA&JCT ANN1}.
Using the Ward identities for high energies we have derived the expression for the quark loop
three-point function with two Coulomb and one transverse line,

\begin{equation}
\label{2}
V_{00i}(k_1, k_2, k_3)\approx {{K_{2i}}\over{k_{10}}}\left[k_{20}S(k_2)+k_{30}S(k_3)\right]
-{{K_{1i}}\over{k_{20}}}\left[k_{30}S(k_3)+k_{10}S(k_1)\right],
\end{equation}

where the gluon self-energy from the quark loop is

\begin{equation}
\label{3}
\mbox{tr}(t^at^b)S_{\mu_1\mu_2}(p)=g^2C_q\delta^{ab}(p_{\mu_1}p_{\mu_2}-p^2\delta_{\mu_1\mu_2})S(p^2)
\end{equation}

with

\begin{equation}
\label{4}
S(p^2)=8i\pi^{2-{{\epsilon}\over 2}}\Gamma({{\epsilon}\over 2}){{\Gamma^2(2-{{\epsilon}\over 2})}\over{\Gamma(4-\epsilon)}}
\left[(-p^2-i\eta)^{-{{\epsilon}\over 2}} -(\mu^2)^{-{{\epsilon}\over 2}}\right],
\end{equation}

where a renormalization subtraction at a mass $ \mu $ has been made and $ \epsilon =4-n $, $n$ is the number of space-time
dimensions. Applying (2), (3) and (4) to the graph in fig.1 we obtain for the $K^2$part the expression

\begin{eqnarray}
\label{5}
2B(b, 0i0)=-{1\over 4}g^6(2\pi)^{-8}C^2_qT(R)\delta_{ab}K^2\int d^4p \int d^4q {{P_iQ'_j}\over{P^2P'^2Q^2Q'^2}} \nonumber \\
\times \{{1\over{p_0p'_0}}[S(r')-S(q)]+{1\over{p_0q_0}}[S(r')-S(p')]-{1\over{p_0r'_0}}[S(q)+S(p')]\}.
\end{eqnarray}

The momenta are defined as $ p'=p-k $, $ q'=q-k$, $r'=k-p-q$, $ p^2=p_0^2-P^2$ and in the high energy limit we have used the
approximation

\begin{equation}
\label{6}
 {{p_0}\over{p_0^2-P^2+i\eta}}\approx {1\over{p_0}}.
\end{equation}

The first term in (5) is explicit linear energy divergence. It is the difference of two integrals, one with $ S(r') $
and the other with $ S(q)$.

\section{Linear energy divergence}

Let us consider the first integral in (5).

\begin{eqnarray}
\label{7}
J_{ij}=\Gamma({{\epsilon}\over 2})\int d^4p \int d^4q {{p_0}\over{p^2_0-P^2+i\eta}}\cdot {{p'_0}\over{p'^2_0-P'^2+i\eta}} \nonumber \\
\times {{P_iQ'_j}\over{P^2P'^2Q^2Q'^2}}\left[(p_0+q_0-k_0)^2-(P+Q-K)^2+i\eta \right]^{-{{\epsilon}\over 2}}
\end{eqnarray}

Using the Schwinger representation for the propagators $ J_{ij} $ becomes

\begin{eqnarray}
\label{8}
J_{ij}=(-i)^{2+{{\epsilon}\over 2}}\int_{-\infty}^{\infty}dp_0 p_0(p-k)_0 \int d^{3-\epsilon}P
  \int _{-\infty}^{\infty}dq_0 \int d^{3-\epsilon} Q \nonumber \\
\times \int_{0}^{\infty} d \alpha \int_{0}^{\infty} d \beta \int_{0}^{\infty} d \gamma {\gamma}^{{{\epsilon}\over 2}-1} {{P_iQ'_j}\over{P^2P'^2Q^2Q'^2}} \nonumber \\
\times e^{i\alpha (p^2_0-P^2+i\eta)} e^{i\beta (p'^2_0-P'^2+i\eta)} e^{i\gamma (r'^2_0-R'^2+i\eta)}
\end{eqnarray}

Performing the $ q_0 $ and $p_0 $ integrations with Gaussian integrals followed by integration over the parameter $ \gamma $,
we obtain

\begin{eqnarray}
\label{9}
J_{ij}=(-i)^{3+{{\epsilon}\over 2}} \pi \Gamma({{\epsilon -1}\over 2})\int d^{3-\epsilon}P\int d^{3-\epsilon} Q 
  {{P_iQ'_j}\over{P^2P'^2Q^2Q'^2}} \int_{0}^{\infty} d \alpha \int_{0}^{\infty} d \beta
   e^{-i\alpha P^2-i\beta P'^2-\eta (\alpha + \beta)} \nonumber \\ 
\times {1\over {(\alpha + \beta)^{3/2}}} e^{ik^2_0{{\alpha \beta}\over{\alpha + \beta}}}
 \cdot \left[ {{i}\over 2} -{{\alpha \beta}\over{\alpha + \beta}}k^2_0\right]
 \cdot (\eta+iR'^2)^{{{1-\epsilon}\over2}}
\end{eqnarray}

 Changing the variables of integration $ \alpha$ and $\beta $ as

\begin{eqnarray}
\label{10}
\alpha=\lambda v, ~~~~~~~~~~\beta=\lambda (1-v),~~~~~~~~~~~~~
({{\partial \alpha, \partial \beta}\over{\partial \lambda, \partial v}})=\lambda,  \nonumber \\
0<v<1, ~~~~~~0<\lambda <\infty,
\end{eqnarray}

makes $\lambda$-integration easy, leading to

\begin{eqnarray}
\label{11}
J_{ij}=\Gamma({{\epsilon-1}\over 2}){1\over 2}\pi^{{3\over 2}}e^{-i\epsilon{{\pi}\over 2}}
\int d^{3-\epsilon} P\int d^{3-\epsilon} Q 
\int_{0}^{1}dv{{P^2v+P'^2(1-v)}\over{[P^2v+P'^2(1-v)-k^2_0v(1-v)-i\eta]^{3/2}}} \nonumber  \\
\times {{P_iQ'_j}\over{P^2P'^2Q^2Q'^2}}
\cdot{1\over{(R'^2)^{{\epsilon-1}\over 2}}}.
\end{eqnarray}

Repeating the same operations with the other integral containing $ S(q) $, we obtain for the linear energy divergence term in (5)
the expression

\begin{eqnarray}
\label{12}
L_{ij}&=&\int d^4p \int d^4 q {{P_iQ'_j}\over{P^2P'^2Q^2Q'^2}}\cdot {1\over{p_0p'_0}}\left[S(r')-S(q)\right ]\nonumber \\
&=&\Gamma({{\epsilon-1}\over 2}){1\over 2}\pi^{3/2}e^{-i\epsilon{{\pi}\over 2}}\int d^{3-\epsilon}P \int d^{3-\epsilon} Q
 \int_{0}^{1} dv{{P^2v+P'^2(1-v)}\over{[P^2v+P'^2(1-v)-k^2_0v(1-v)-i\eta]^{3/2}}} \nonumber \\
&&\times {{P_iQ'_j}\over{P^2P'^2Q^2Q'^2}}\left[{1\over{(R'^2)^{{\epsilon-1}\over 2}}} -{1\over{({Q'}^2)^{{\epsilon-1}\over 2}}}\right].
\end{eqnarray}

The linear energy divergence reflects in the factor $ \Gamma({{\epsilon-1}\over 2}) $. The $ Q$- integral is

\begin{equation}
\label{13}
X_j=\int d^{3-\epsilon} Q{{Q'_j}\over{Q^2Q'^2}}
\left[{1\over{(R'^2)^{{\epsilon-1}\over 2}}}-{1\over{(Q'^2)^{{\epsilon-1}\over 2}}}\right].
\end{equation}

The second integral is easy.

\begin{eqnarray}
\label{14}
B_j=\int d^{3-\epsilon}Q{{Q'_j}\over{Q^2(Q'^2)^{{{1+\epsilon}\over 2}}}} \nonumber \\
=-K_j(K^2)^{-\epsilon}\pi^{{{3-\epsilon}\over 2}}{{\Gamma(\epsilon)}\over{\Gamma({{1+\epsilon}\over 2})}}
\cdot{{\Gamma(2-\epsilon)\Gamma({{1-\epsilon}\over 2})}\over{\Gamma({5\over 2}-{{3\epsilon}\over 2})}}
\end{eqnarray}

Let us study the first integral in (13).

\begin{equation}
\label{15}
A_j=\int d^{3-\epsilon} Q{{Q'_j}\over{Q^2Q'^2}}\cdot{1\over{(R'^2)^{{{\epsilon-1}\over 2}}}}
\end{equation}

We combine the denominators $Q^2$ and $Q'^2$ with the Feynman parameter $x$ and then $(R'^2)^{{{\epsilon-1}\over 2}} $
with the parameter $y$ (remembering that $R'=-P'-Q$, $Q'=Q-K$), then integrate in $ d^{3-\epsilon}Q$.

\begin{eqnarray}
\label{16}
A_j=\pi^{{{3-\epsilon}\over 2}}{{\Gamma(\epsilon)}\over{\Gamma({{\epsilon-1}\over 2})}}
\int_{0}^{1}dx\int_{0}^{1}dy y^{{{\epsilon-3}\over 2}} (1-y)
\cdot[K_jx(1-y)-K_j-P'_jy] \nonumber \\
\times \{P'^2y+K^2x(1-y)-[Kx(1-y)-P'y]^2\}^{-\epsilon}
\end{eqnarray}

We insert (16) and (14) into (12).

\begin{eqnarray}
\label{17}
L_{ij}={1\over 2}\pi^{3-{{\epsilon}\over 2}}e^{-i\epsilon{{\pi}\over 2}} \int d^{3-\epsilon}P
\int_{0}^{1}dv{{P^2v+P'^2(1-v)}\over{[P^2v+P'^2(1-v)-k^2_0v(1-v)-i\eta]^{{3\over 2}}}}\cdot{{P_i}\over{P^2P'^2}} \nonumber \\
\times \{\Gamma(\epsilon)\int_{0}^{1}dx\int_{0}^{1}dy y^{{{\epsilon-3}\over 2}}(1-y)[K_jx(1-y)-K_j-P'_jy]
\cdot[P'^2y+K^2x(1-y)-(Kx(1-y)-P'y)^2]^{-\epsilon} \nonumber \\
+\Gamma({{\epsilon-1}\over 2})\Gamma(\epsilon)
\cdot{{\Gamma(2-\epsilon)\Gamma({{1-\epsilon}\over 2})}\over{\Gamma({{1+\epsilon}\over 2})\Gamma({5\over 2}-{{3\epsilon}\over 2})}} 
\cdot K_j(K^2)^{-\epsilon}\}
\end{eqnarray}

We notice that the $d^{3-\epsilon}P$ integration is IR safe and also UV finite by power counting for $K_j$ terms, while it is
UV divergent for $P'_jy$ term. Hence, for the leading divergence we can set

\begin{equation}\label{18}
[P'^2y+K^2x(1-y)-(Kx(1-y)-P'y)^2]^{-\epsilon} \approx 1,
\end{equation}

\begin{eqnarray}
\label{19}
L_{ij}=\Gamma({{\epsilon-1}\over 2})\Gamma(\epsilon){1\over 2}\pi^3K_j\int d^{3-\epsilon}P\int_{0}^{1}dv
{{P_i}\over{P^2P'^2}}\cdot{{P^2v+P'^2(1-v)}\over{[P^2v+P'^2(1-v)-k^2_0v(1-v)-i\eta]^{3\over 2}}} \nonumber \\
\times \{ {{\Gamma(3)}\over{2\Gamma({{\epsilon+5}\over 2})}} -{{\Gamma(2)}\over{\Gamma({{\epsilon+3}\over 2})}}
+{{\Gamma(2-\epsilon)\Gamma({{1-\epsilon}\over 2})}\over{\Gamma({{1+\epsilon}\over 2})\Gamma({5\over 2}-{{3\epsilon}\over 2})}}\}
\nonumber \\
-{1\over 2}\pi^3\Gamma(\epsilon)\int d^{3-\epsilon}P\int_{0}^{1}dv{{P^2v+P'^2(1-v)}\over{[P^2v+P'^2(1-v)-k^2_0v(1-v)-i\eta]^{3/2}}}
\cdot{{P_iP'_j}\over{P^2P'^2}}\int_{0}^{1}dx \int_{0}^{1}dy y^{{{\epsilon-1}\over 2}}(1-y).
\end{eqnarray}

We can easily isolate the UV divergence from the last integral in (19). It  behaves like $ \Gamma({{\epsilon}\over 2})$.

\section{Conclusion}

Individual Feynman graphs in Coulomb gauge QCD to three loop order contain even linear energy divergences.
Our analysis shows their behaviour. It is

\begin{equation}
\label{20}
L_{ij}=\Gamma({{\epsilon-1}\over 2})\Gamma(\epsilon)K_iK_jf(k^2_0, K^2)
+\Gamma(\epsilon)\Gamma({{\epsilon}\over 2})K_iK_jg(k^2_0, K^2).
\end{equation}

The first term is the product of poles in $ {1\over{\epsilon-1}}\cdot {1\over{\epsilon}}$ while the second is
 a double pole in $ {1\over{\epsilon^2}}$. Such dangerous divergences do cancel in the sum of graphs with three-point and
four-point fermion loop insertions, but next to leading divergences coming from ${1\over{p_0q_0}}$, ${1\over{p_0r'_0}}$
and ${1\over{q_0r'_0}}$ terms do not. Hence, UV divergences from higher order graphs cannot be consistently absorbed by
renormalization of the Christ-Lee term \cite{AA&JCT ANN2}.

\end{document}